\documentclass{nature}

\usepackage{graphicx}
\usepackage{amsmath}
\usepackage{amssymb}
\usepackage{color}
\usepackage{dcolumn}
\usepackage{epsfig}
\usepackage{bm}
\usepackage[urlcolor=blue]{hyperref}
\hypersetup{backref, colorlinks=true, linkcolor=blue, citecolor=blue}

\title{Topological superconductivity in a topological insulator}

\author{Hao Yu$^{1,2,5}$, Noah F. Q. Yuan$^{1,5}$, Wei-Jian Li$^{3,2}$, Liu-Cheng Chen$^{1,2}$, Zi-Yu Cao$^{1}$, Di Li$^{4}$, Xiaoying Qin$^{4}$, Chang-De Gong$^{3}$ \& Xiao-Jia Chen$^{1,2\ast}$}

\begin{document}

\maketitle

\begin{affiliations}
\item{School of Science, Harbin Institute of Technology, Shenzhen 518055, China}
\item{Center for High Pressure Science and Technology Advanced Research, Shanghai, 201203, China}
\item{National Laboratory of Solid State Microstructures and School of Physics, Nanjing University, Nanjing 210093, China}
\item{Key Laboratory of Materials Physics, Institute of Solid State Physics, Chinese Academy of Sciences, Hefei 230031, China}
\item{These authors contribute equally to this work.}\\
$^{\ast}$E-mail: xjchen2@gmail.com
\end{affiliations}

\begin{abstract}
Topological superconductivity as an exotic quantum phenomenon with coupled nontrivial topological order and superconductivity together in a single substance has drawn extensive attention because of its novelty as well as the potential for quantum computation\cite{Kitaev,kitaev,nayak,read-green,ivanov,Alicea}. A direct idea for producing topological superconductors is to create superconductivity based on the well recognized topological insulators\cite{TKNN,QSH1,QSH2,TI1,TI2,TI3,hasa,xlqi}. The topological insulating states in highly efficient thermoelectric materials Bi$_2$Te$_3$ and Bi$_2$Se$_3$ and their alloy Bi$_{2}$Te$_{3-x}$Se$_{x}$ have been established from angle-resolved photoemission\cite{ylch,yxia,cych} and transport\cite{dxqu} experiments. Superconductivity was also observed based on these popular topological insulators by the application of pressure\cite{Ilin,jlzh,kirs}, chemical dopant\cite{yhor}, and heterostructures\cite{jpxu,qlhe}. However, the experiments mainly focusing on Bi$_{2}$Se$_3$ doped by metals such as Cu\cite{wray,krie,levy,mata,yone,chu,ful1,ful2,ful3,lhao,hsie}, Sr\cite{Sr1,Sr2,Sr3,Sr4}, and Nb\cite{Nb0,Nb1,Nb2,Nb3,Nb4,Nb5,Nb6} have not provided the consistent evidence to support the topological superconductivity. 
Here we carry out a systematic high-pressure study on a topological insulator Bi$_{2}$Te$_{2.7}$Se$_{0.3}$ to provide the convincing evidence for the expected topological superconductivity. Four phases with different structures are found upon compression by the X-ray diffraction, Raman scattering, and electrical transport measurements. The Hall resistivity and electronic structure calculations help to identify the topological surface state in the entire initial phase, while superconductivity is found to coexist with such a state of the compressed material after its passing the electronic topological transition, followed by three other superconducting phases without topological character. For these superconducting phases, we observe that the upper critical field follows with the temperature in the critical exponent ${2/3}$ for the first one with the topological surface state and $1$ for the left. These observations support the realization of the topological superconductivity in the initial phase according to the theoretically proposed critical field measure. 
Therefore, this work points out a big pool and new direction for finding topological superconductors from topological thermoelectric materials. The results also demonstrate that the critical field measure is an easy and powerful method for the identification of topological superconductivity in the system where topological surface states and superconductivity coexist. 
\end{abstract}

Symmetry and topology are two of the main themes of modern physics, especially the condensed matter physics in recent decades. In two dimensions (2D), quantum (anomalous) Hall effects emerge with nontrivial electronic topology when the time-reversal symmetry is broken\cite{TKNN}. When the time-reversal symmetry is preserved, quantum Hall states then develop into the quantum spin Hall (QSH) phase\cite{QSH1,QSH2}. In three dimensions (3D), stacked QSH states evolve into topological insulators (TIs)\cite{TI1,TI2,TI3}. The crystal symmetries further intertwine with the electronic topology, leading to topological crystalline insulators\cite{TCI,TCI1,TCI2}. 
These topological insulating phases have been experimentally verified with high precision in several materials, in particular highly efficient thermoelectric materials Bi$_2$Te$_3$ and Bi$_2$Se$_3$ and their alloy Bi$_{2}$Te$_{3-x}$Se$_{x}$\cite{TI3,hasa,xlqi,ylch,yxia,cych,dxqu}. 
Importantly, such TI materials have also been found to be intrinsically superconducting by the application of pressure\cite{Ilin,jlzh,kirs} or chemical dopant\cite{yhor,Sr1,Sr2,Nb0,Nb1}, which naturally invites two questions on the nature of this kind of superconductivity. From the symmetry wise, one needs to determine the pairing symmetry of the intrinsic superconductivity. From the topology wise, one needs to know whether and how the topological superconductivity can be realized in these topologically nontrivial materials.

Theoretical proposals followed by various experimental measurements have been carried out to address these two questions. 
Classified by the irreducible representations of the point group $D_{3d}$, the possible pairing symmetry of Bi(Te,Se) systems has been theoretically proposed as $A_{1g}$ phase ($s$-wave), odd-parity $A_{1u}$ phase\cite{ful1,lhao,hsie} or two-component $E_u$ phase (nematic\cite{ful2,ful3} versus chiral\cite{Nb2}). For doped topological insulators, some experimental data support the $A_{1g}$ phase ($s$-wave)\cite{levy} without topological superconductivity\cite{chu} for Cu-doped Bi$_2$Se$_3$, while others support the nematic $E_u$ phase for Bi$_2$Se$_3$ doped with Cu\cite{krie,mata,yone}, Sr\cite{Sr1,Sr2,Sr3,Sr4} or Nb\cite{Nb1,Nb3,Nb4,Nb5,Nb6}, which are believed to host topological superconductivity, although discrepancy still remains\cite{Nb6,ful3}.
For topological insulators under pressure, experimental data indicate unconventional superconductivity in Bi$_2$Se$_3$\cite{kirs}, while no direct evidence on unconventional superconductivity has been provided in Bi$_2$Te$_3$\cite{jlzh}. The issue regarding whether topological superconductivity can take place in topological insulators has not been settled yet but is highly desired from the science community. To address this issue, we conduct a careful experimental and theoretical study on both the normal phase and superconducting phase of Bi$_{2}$Te$_{2.7}$Se$_{0.3}$ at high pressures. The details are given as follows. 

The room-temperature phase and structure behaviours of Bi$_{2}$Te$_{2.7}$Se$_{0.3}$ at ambient pressure and high pressures were determined by X-ray diffraction measurements. The details for the experiments and refinements are given in Methods. The diffraction patterns and profiles indicate the occurrence of several phases at high pressures (Fig. 1\textbf{a}). The Rietveld refinements for four representative pressures yield good agreement between the measurements and calculations (Fig. 1\textbf{b}). The obtained structural parameters are summarized in Extended Data Table 1. The results support the formation of three new phases besides the initial one. All the obtained phases are named as I to IV with space groups of $R$-3$m$, $C$2/$m$, $C$2/$c$, and $Im$-3$m$ from the ambient to high pressure, respectively. The evolution of the volume per formula of these phases and the atomic arrangements for Bi and Te(Se) in each structure are presented in Fig. 1\textbf{c}. Phase I can maintain its initial rhombohedra structure up to 10 GPa. For Phase IV, all three kinds of atoms could occupy the same position with equal possibility and are hence geometrically indistinguishable. The obtained phases and structures of Bi$_{2}$Te$_{2.7}$Se$_{0.3}$ are consistent with the previous studies on the two ending compounds Bi$_{2}$Te$_{3}$\cite{lzhu} and Bi$_{2}$Se$_{3}$\cite{vila}. Interestingly, the axial ratio $c/a$ of the lattice parameters exhibits non-monotonic behaviour under pressure with a minimum at around 2 GPa (Fig. 1\textbf{d}). Such a minimum was also observed in both Bi$_{2}$Te$_{3}$\cite{naka} and Bi$_{2}$Se$_{3}$\cite{vila}. It has been proposed to be associated with the Lifshitz transition, also known as the electronic topological transition (ETT)\cite{naka}. The experimental evidence for pressure-induced ETT in Bi$_{2}$(Te,Se)$_{3}$-based system through various techniques is summarized in Methods. For $n$-type Bi$_{2}$Te$_{2.7}$Se$_{0.3}$, the ETT near 2 GPa has been firmly confirmed through the pressure dependence of $\sigma$, $S$, their power factor $\sigma S^{2}$, effective mass, and charge mobility\cite{koro}. 

The structural characters of the obtained four phases were further examined through high-pressure Raman scattering measurements. The phonon modes appear in accordance with the structural symmetry of each phase (Extended Data Figs. 1\textbf{a} and 1\textbf{b}). For example, the $R$-3$m$ symmetry predicts four phonon modes for the first phase (I)\cite{rich}. We do observe these modes in the pressure range up to 8 GPa. The pressure dependence of the low-frequency $E_{g}^{1}$ mode is the first report in history\cite{vila,rich}. The flat background at high pressures is just the feature of Phase IV with space group of $Im$-3$m$ because no active Raman modes are expected from the symmetry consideration. The details for the mode assignment are given in Methods. The pressure dependence of the frequencies of the obtained phonon modes provide the accurate boundaries between phases and their coexistence (Extended Data Fig. 1\textbf{c}). Note that the slope change of the pressure-dependent frequency can be observed at around 2 GPa for all the phonon modes in Phase I. Meanwhile, the full width at half maximum (FWHM) has the maximum (minimum) for the $E_{g}^{1}$ ($E_{g}^{2}$) mode at pressure near 2 GPa where the FWHM of the $A_{1g}^{1}$ and $A_{1g}^{2}$ mode changes slope (Extended Data Fig. 1\textbf{d}). These offer additional spectroscopic evidence for the ETT in Phase I of the studied material.

Hall effect measurements at low temperature of 10 K were carried out to examine the validity of the nontrivial topological surface states under pressure. The Hall resistivity $\rho_{xy}$ vs magnetic field curves for representative pressures reveal the different nature of the carriers in the obtained four phases (Fig. 2\textbf{a}). The first phase (I) and the last two phases (III and IV) hold the $n$-type feature, while Phase II becomes a $p$-type material. This indicates that the electrical transport properties of the two intermediate phases (II and III) are controlled by the different types of the carriers. The robust topological order in Bi$_{2}$Te$_{3-x}$Se$_{x}$ has been reported from the angle-resolved photoemission measurements\cite{cych}. The Hall conductivity data of Phase I at various pressures exhibit a nonlinear resonant structure when applying magnetic fields (Extended Data Fig. 2), evidencing the existence of the surface state in topological insulators\cite{dxqu}. The Hall conductivity can be decomposed into the surface and bulk components based on the method developed recently\cite{dxqu}. The details for determining the mobility of the surface $\mu_s$ and bulk $\mu_b$ state as well as the carrier concentration $n$ of the bulk are described in Methods. Fitting our experimental data yields the Hall conductivity for the surface $\sigma_{xy}^s$ and the bulk $\sigma_{xy}^b$, supporting the existence of the nontrivial topological states in Phase I (Fig. 2\textbf{b} and Extended Data Fig. 2). However, the topological states does not survive in the rest three high-pressure phases reflected from the absence of the resonant structure of the Hall conductivity.   
       
The obtained surface mobility $\mu_{s}$ is one order larger in magnitude than the bulk value $\mu_{b}$ over the whole pressure region studied in Phase I, supporting the nontrivial topological state. As can be seen from Fig. 2\textbf{c}, both mobilities increase with increasing pressure. The bulk one reaches a maximum at the boundary between Phase I and II. The mobility decreases with increasing pressure in Phase II. There is a minimum at the boundary between Phase III and IV. The distinct phase boundaries can be clearly established from the pressure dependence of the effective carrier concentration $n_{eff}$ for the bulk state in Phase I and the Hall concentration $n_{H}$ in the rest three phases (Extended Data Fig. 3). These boundaries seemingly do not shift with temperature when keeping pressure unchanged. Now the phases and their boundaries are well determined for Bi$_{2}$Te$_{2.7}$Se$_{0.3}$ from X-ray diffraction, Raman scattering, and Hall effect measurements. 

We perform the first-principles calculations of the Bi$_{2}$Te$_{2.7}$Se$_{0.3}$ alloy in the $R$-3$m$ phase under pressures (Fig. 3\textbf{a}). For the chosen first Brillouin zone and the projection along the (001) surface (Fig. 3\textbf{b}), the metallic surface state appears in all the pressure range studied up to 8 GPa (Fig. 3\textbf{c}), in good agreement with the transport measurements at temperature of 10 K (Fig. 2\textbf{b}). Furthermore, the evolution of energy contours of the top bulk valance band with increasing pressure (Fig. 3\textbf{d}) indicates the ETT around 2 GPa, where energy contours merge together and the number of distinct contours at the same energy changes from 3 to 2 as pressure increases. The numerical result of ETT is consistent with the transport and Raman scattering measurements.

Among the obtained four phases, only Phase I carries on the topological states at room temperature. Therefore, topological superconductivity can be only expected in this phase. To verify this expectation, we then cool down the material and perform transport measurements.

As expected, superconducting transitions are observed at low temperatures with a sudden resistivity drop at 4.3 GPa and even to zero value at 5.5 GPa, and the zero-resistivity states are always achievable for pressures higher than 5.5 GPa (Fig. 4\textbf{a}) within the experimentally achievable pressure range. The suppression of superconductivity by the applied magnetic fields is demonstrated in Fig. 4\textbf{b} for selected pressures of interest. The superconducting phase diagram in the pressure-temperature plane is presented in Fig. 4\textbf{c}. A comparison of the present results with the other experiments is given in Methods.

The same spectroscopic features at room temperature (Extended Data Fig. 1), low temperature of 3.6 K (Extended Data Fig. 4), and the temperatures across $T_{c}$ (Extended Data Fig. 6) indicate that the compound keeps the same electronic phases at low temperatures as determined at room temperature. The nontrivial topological states can coexist with the superconducting state in Phase I at pressures up to 10 GPa (Extended Data Fig. 4). This topological state maintains when the material enters the superconducting state (below $T_{c}$) (Extended Data Fig. 6). In other words, among the obtained four superconducting phases, the superconducting state in Phase I is the only one to coexist with the nontrivial topological states in the studied compound. Therefore, the region at pressures of 5.5-10 GPa and temperatures less than $T_{c}$ in Phase I provides the only window to examine the possibility of topological superconductivity in Bi$_{2}$Te$_{2.7}$Se$_{0.3}$.  

Now we turn to the identification of the coexistence between superconductivity and topological surface states by investigation of critical fields in the Ginzburg-Landau regime near $T_c$. 
In Fig. 4\textbf{b}, two distinct types of critical fields have been obtained. In the three high-pressure phases (II, III, IV) without the topological surface states, the linear temperature dependence $H_{c2}\propto 1-T/T_c$ is found, while in Phase I it can be found that $H_{c2}\sim(1-T/T_{c})^{2/3}$ near $T_{c}$ (Inset of Fig. 4\textbf{b} and Extended Data Fig. 5). This sharp contrast is in excellent agreement with the theoretical prediction of the critical fields in trivial versus topological superconductors. To understand this mechanism, we need to work out the order parameter and the critical field by minimizing Ginzburg-Landau free energy, as elaborated in Ref.\cite{yuan}. In the following we stretch the outline of the theory.
%For this purpose, we define the field direction as the $z$-axis, and the order parameter is $\psi=\psi(\bm r)$ with $\bm r=(x,y)$. %The superconductor is simplified as a semi-infinite region with one open surface.

In the trivial superconductors without surface states, the critical field is linear in temperature.
%The critical field is thus the upper critical field $H_{c2}=\Phi_0/(2\pi\xi^2)\propto 1-T/T_c$ with the flux quantum $\Phi_0=h/(2e)$. If $x=0$ is the open surface and $x\geq 0$ is the superconductor, then 
Denote $\xi\propto(1-T/T_c)^{-1/2}$ as the coherence length of the superconductor, then superconductivity is killed when a flux quantum $\Phi_0=h/(2e)$ is trapped in the area $\sim\xi^2$. 
As a result, $H_{c2}\propto 1/\xi^2\propto 1-T/T_c$, which corresponds to the high-pressure cases as shown for 14.0 GPa and 16.8 GPa (Fig. 4\textbf{b} and Extended Data Fig. 5).

In the topological superconductors with surface states, the critical field can have critical exponent of 2/3 in the temperature dependence. 
%If $z=0$ is the open surface and $z\geq 0$ is the superconductor, then $\psi=\exp(-\frac{1}{4}|\bm r-\bm r_0|^2/\xi^2-z/l_s)$ with the new length scale $l_s$ known as the extrapolation length, which measures the coupling strength between topological states and superconductivity on the surface. The critical field is again the upper critical field $H_{c2}=\Phi_0/(2\pi\xi^2)\propto 1-T/T_c$. 
After the bulk superconductivity is killed, as we further increase magnetic field strength, the surface layer of the thickness $\sim l_s$ prefers to remain superconducting. 
The new length scale $l_s$, known as the extrapolation length, measures the coupling strength between the topological states and superconductivity on the surface. %, which renders the topological surface superconductivity more localized than the trivial one. 
%Near the open surface, the non-Gaussian profile of the order parameter reads $\psi=\exp[-x/l_s-\frac{2}{3}x^3/(\xi^2l_s)]$ with $x$ being the distance from the open surface. 
When a flux quantum is trapped in the area $\sim\zeta^2$ near the surface, superconductivity is killed, where $\zeta=(l_s\xi^2)^{1/3}$ combines the extrapolation length of the surface superconductivity and the coherence length of the bulk superconductivity.
%We can then derive $H_{c2}\sim\Phi_0/\zeta^2$, where $\zeta=(\xi^2l_s)^{1/3}\propto(1-T/T_{c})^{-1/3}$ measures the size of non-Gaussian region of the surface superconductivity.
We then derive $H_{c2}\sim\Phi_0/\zeta^2\sim (1-T/T_c)^{2/3}$, which corresponds to the 6.4 GPa case of Fig. 4\textbf{b} (also see Extended Data Fig. 5).

From the obtained experimental evidence for surface states, we believe the superconducting state of the Bi$_{2}$Te$_{2.7}$Se$_{0.3}$ alloy in Phase I can have nontrivial topology.
When the bulk pairing is $A_{1g}$ phase ($s$-wave), the bulk is fully gapped, surface states form the Fu-Kane model\cite{FuKane} in 2D, and Majorana zero modes should be found in the vortex cores on the surfaces.
When the bulk pairing is $A_{1u}$ phase, the fully-gapped bulk realizes topological superconductivity in class DIII in 3D\cite{classification}, and the the surface states are recognized as helical Majorana surface states, which should lead to an asymmetric surface tunneling spectroscopy\cite{ful1,lhao,hsie}.
The bulk pairing symmetry can also be two-component $E_u$ phase according to recent data, which can spontaneously break in-plane rotation (time-reversal) symmetry and falls into the nematic (chiral) phase. 
In the nematic phase, topological superconductivity in class DIII is realized in 3D, and the surface states are helical Majorana surface states\cite{ful2,ful3}, where the in-plane critical fields should be anisotropic\cite{ful3,Nb6}. 
While in the chiral phase, bulk superconductivity is nodal with point nodes, and Weyl superconductivity is realized. The surface states are thus Majorana arc states connecting surface-projected Weyl points with opposite Chern numbers\cite{Nb2}. In the chiral phase, time-reversal breaking signals such as Kerr rotation should be observed. 
These results and discussions above are summarized in Table 1.
%More importantly, our new method of critical field measurement for judging topological superconductivity avoids the discussion on microscopic details of the superconductor, and hence can be applied across a wide range of cases. 

The search for topological superconductivity in topological materials such as topological insulators has been longed for in condensed matter physics community during the recent decade. In this work, from carefully designed experiments, and detailed electrical transport, X-ray diffraction, and Raman scattering measurements, we first narrow down the coexisting window of the topological surface states and superconductivity for the same compound. We then apply the detection method proposed recently to measure the critical exponent of critical fields in the temperature dependence, regardless of microscopic details. Finally, we find both the topologically nontrivial and trivial cases for the different superconducting phases predicted by the theory, and hence besides evidence for topological superconductivity, at the same time we also validate this recently proposed method for identifying and distinguishing topological and conventional superconductivity. 
The current work thus realizes the long-standing dream for the topological superconductivity in most popular topological insulators Bi$_2$Se$_{3-x}$Te$_{x}$ $-$ the alloy compounds of Bi$_2$Se$_3$ and Bi$_2$Te$_3$.

\newpage

\vspace{0.3cm}
\noindent\textbf{Extended Data Table 1 $\mid$ Detailed structural parameters of Bi$_{2}$Te$_{2.7}$Se$_{0.3}$ at room temperature}. 
\begin{center}
\vspace{-1cm}
\begin{tabular}{c c c c c c c c c c }
  \hline
  \hline
  P (GPa) &Space group & Lattice parameters &  Atom & Site & $x$ & $y$ & $z$    \\
    \hline
  {1.0} &$R$-3$m$& $a$=4.391(3) \AA\/ & Bi  & 6$c$ & 0 &  0  & 0.4080(1)\\
  && $c$=30.482(27) \AA\/ &  Te/Se(1)& 6$c$ & 0  & 0 & 0.3524(1)\\
  &&$R_p$=1.6\%, $R_{wp}$=2.2\% & Te/Se(2)& 3$a$ & 0  & 0 & 0 \\
   &&&  &  &   & & \\
 
 {13.3} &$C$2/$m$& $a$=14.624(20) \AA\/ & Bi(1)  & 4$i$ & 0.1654(3) &  0     &0.1903(7)\\
  && $b$=4.032(4) \AA\/ &  Bi(2) & 4$i$ & 0.4498(9)  & 0     &0.2323(2)\\
  && $c$=17.243(19) \AA\/ & Te/Se(1)& 4$i$ & 0.2356(5)  & 0     &0.4047(8) \\
  &&$\beta$=148.229(21) $^\circ$& Te/Se(2)& 4$i$ & 0.0098(6)  & 0     &0.5745(7) \\
  && $R_p$=1.7\%, $R_{wp}$=2.3\% & Te/Se(3) & 4$i$ & 0.3352(1)  & 0     &0.9808(8) \\
   &&&  &  &   & & \\
 
 {20.6} &$C$2/$c$& $a$=10.145(29) \AA\/ & {Bi} & 8$f$ & {0.2586(6)} & {0.1155(4)}&{0.8333(4)}\\
  && $b$=6.728(13)  \AA\/ &Te/Se(1) & 8$f$ & 0.6013(5) &  0.3709(9)  & 0.9777(5) \\
  && $c$=10.467(26)  \AA\/ &Te/Se(2) & 4$e$ & 0  & 0.5946(9) & 0.25 \\
  && $\beta$=134.373(39) $^\circ$ &  &  &   & & \\
   && $R_p$=7.5\%, $R_{wp}$=8.8\% &  &  &   & & \\
   &&&  &  &   & & \\
 
 {28.0} &$Im$-3$m$& $a$=3.611(16) \AA\/ & Bi  & 2$a$ & 0 &  0 & 0 \\
  &&$R_p$=7.2\%, $R_{wp}$=7.4\%&Te/Se& 2$a$ & 0  & 0 &0\\
  \hline
  \hline
\end{tabular}
\end{center}

\newpage
\begin{center}
\includegraphics[width=\columnwidth]{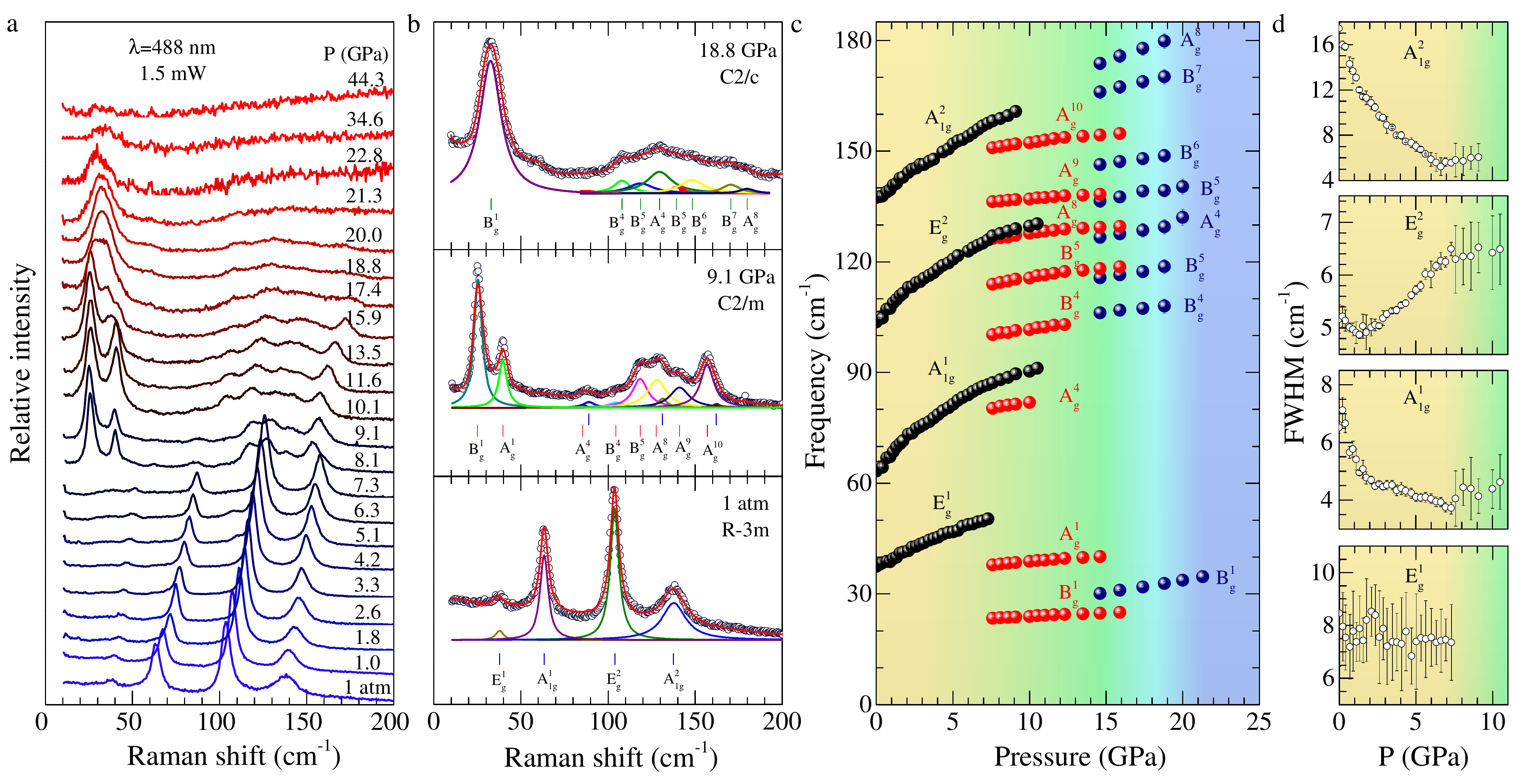}
\end{center}
\vspace{-0.5cm}
\noindent\textbf{Extended Data Fig. 1 $\mid$ High-pressure vibrational properties of Bi$_{2}$Te$_{2.7}$Se$_{0.3}$ at room temperature}. \textbf{a}, Raman scattering spectra collected at various pressures up to 44.3 GPa at room temperature with the laser excitation of 488 nm and laser power of 1.5 mW. The spectra change profiles at pressures of 8.1 and 17.4 GPa. No active Raman peaks at higher pressures above 22.8 GPa are observed. \textbf{b}, From the bottom to top, Representative Raman spectra in open cycles for the phase with space group of $R$-3$m$ at ambient pressure, $C$2/$m$ at 9.2 GPa, and $C$2/$c$ at18.2 GPa, respectively. Lorentz fitting to the data points is shown in the curve for individual phonon mode and their combination. The sticks at each bottom denote the calculated frequencies of the Raman-active modes for the corresponding structure symmetry. \textbf{c}, Pressure dependence of the frequency of each phonon mode in the first three phases with the space group of $R$-3$m$ (Phase I), $C$2/$m$ (Phase II), and $C$2/$c$ (Phase III), respectively. The regime for each phase, the boundary in between of them, and their coexistence can be judged from the data points. \textbf{d}, Pressure dependence of FWHM of the phonon mode $E_{g}^{1}$, $A_{1g}^{1}$, $E_{g}^{2}$, and $A_{1g}^{2}$ (from the bottom to top) for the Phase I with the space group of $R$-3$m$. There is a maximum (minimum) for the $E_{g}^{1}$ ($E_{g}^{2}$) mode at a pressure close to 2 GPa at which the $A_{1g}^{1}$ and $A_{1g}^{2}$ mode changes curvature.  

\newpage
\begin{center}
\includegraphics[width=\columnwidth]{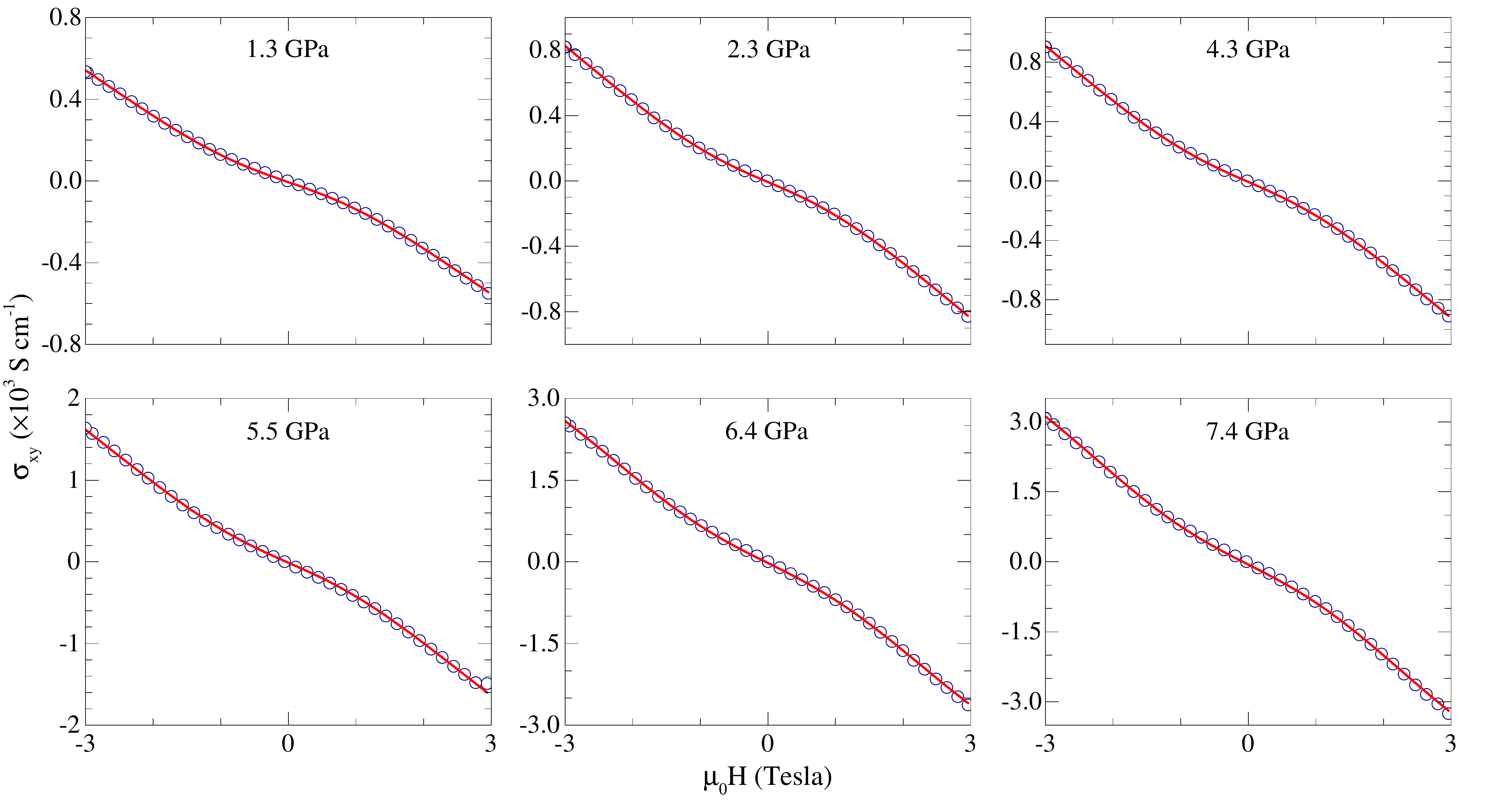}
\end{center}
\vspace{-0.5cm}
\noindent\textbf{Extended Data Fig. 2 $\mid$ Hall conductivity $\sigma_{xy}$ vs applied magnetic field for Bi$_{2}$Te$_{2.7}$Se$_{0.3}$}. The data were taken for Phase I with the space group of $R$-3$m$ at selected pressures and at temperature of 10 K. The open cycles are the experimental data points. The curves are the fitting results to the experiments by considering the combination effects from the bulk and surface states.

\newpage
\begin{center}
\includegraphics[width=\columnwidth]{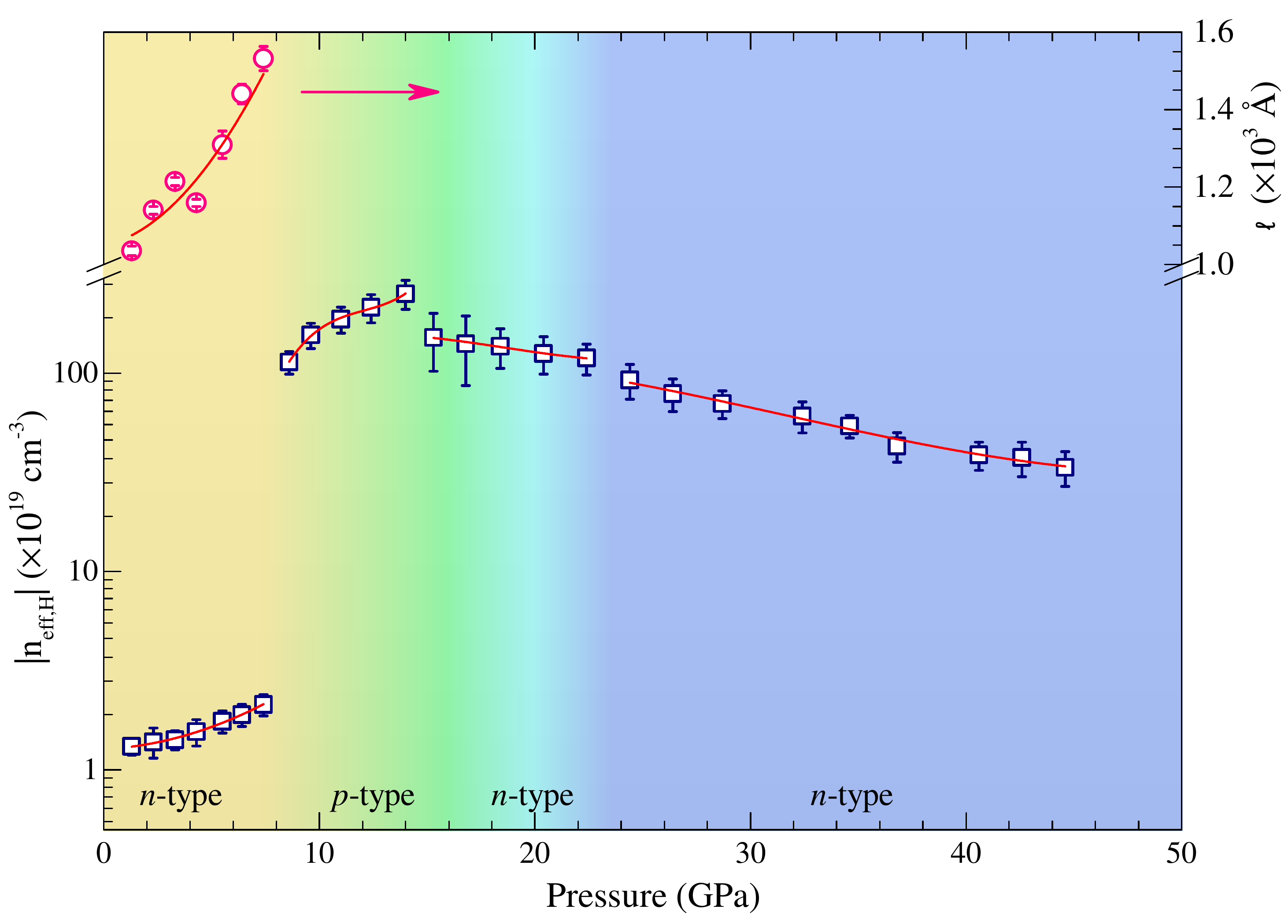}
\end{center}
\vspace{-0.5cm}
\noindent\textbf{Extended Data Fig. 3 $\mid$ Pressure dependence of the bulk carrier concentration in the absolute value for Bi$_{2}$Te$_{2.7}$Se$_{0.3}$ at temperature of 10 K}. Left panel: The effective carrier concentration $n_{eff}$ of the bulk in Phase I and the bulk carrier concentration $n_{H}$ of the rest phases. Right panel: Mean free path $l$ of Phase I. The line is drawn for the guide to the eye. The material is $n$-type in Phase I, III, IV but $p$ type in Phase II. This character can be judged from Fig. 2\textbf{a}.

\newpage
\begin{center}
\includegraphics[width=\columnwidth]{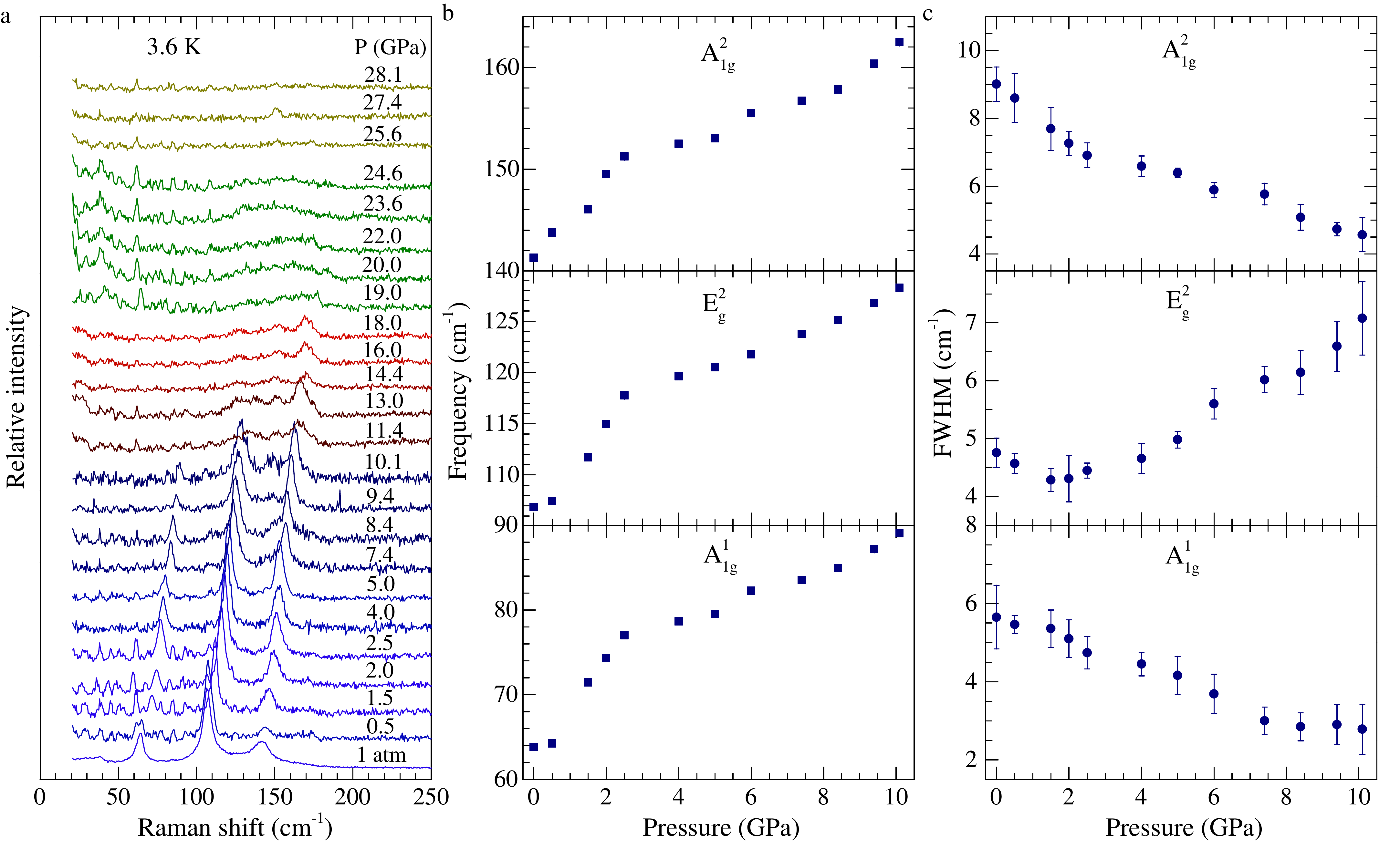}
\end{center}
\vspace{-0.5cm}
\noindent\textbf{Extended Data Fig. 4 $\mid$ Evolution of the phonon modes with pressure for Bi$_{2}$Te$_{2.7}$Se$_{0.3}$ at temperature of 3.6 K}. \textbf{a}, Low-temperature Raman spectra at various pressures up to 28.1 GPa. Four phases can be identified from the different profiles of their phonon modes. In the superconducting state of each phase, the Raman spectra have the same profiles compared with the normal state at the room temperature (Extended Data Fig. 1). This indicates that the superconducting phase carries on all the features of this phase at the normal state. The first superconducting phase thus keeps its nontrivial topological order. \textbf{b} and \textbf{c}, Pressure dependence of the frequency and FWHM of the $A_{1g}^{1}$, $E_{g}^{2}$, and $A_{1g}^{2}$ mode in Phase I with the space group of $R$-3$m$.  

\newpage
\begin{center}
\includegraphics[width=0.7\columnwidth]{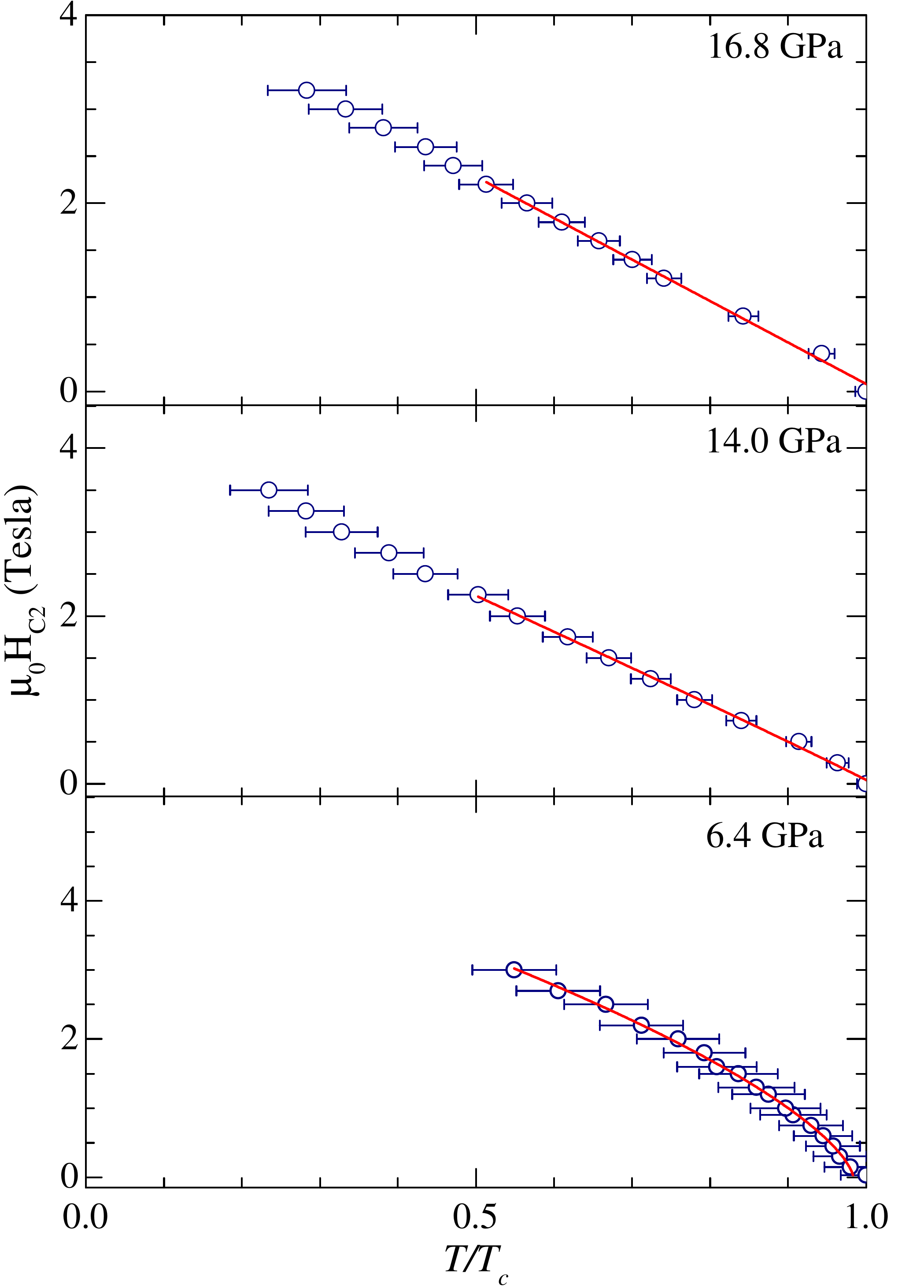}
\end{center}
\vspace{-0.5cm}
\noindent\textbf{Extended Data Fig. 5 $\mid$ Enlarged view of the upper critical field as a function of the reduced temperature ($T/T_{c}$) of Bi$_{2}$Te$_{2.7}$Se$_{0.3}$ at selected pressures}. The experimental data are represented by open cycles. The fitting to the experimental data points near $T_{c}$ in the $T/T_{c}$ range of 0.5 to 1 is shown in the curve by using $\mu_{0}H_{c2}\sim (1-T/T_{c})^{\alpha}$ with $\alpha$ of $2/3$ for pressure of 6.4 GPa and of 1.0 for pressure of 14.0 and 16.8 GPa, respectively.

\newpage
\begin{center}
\includegraphics[width=\columnwidth]{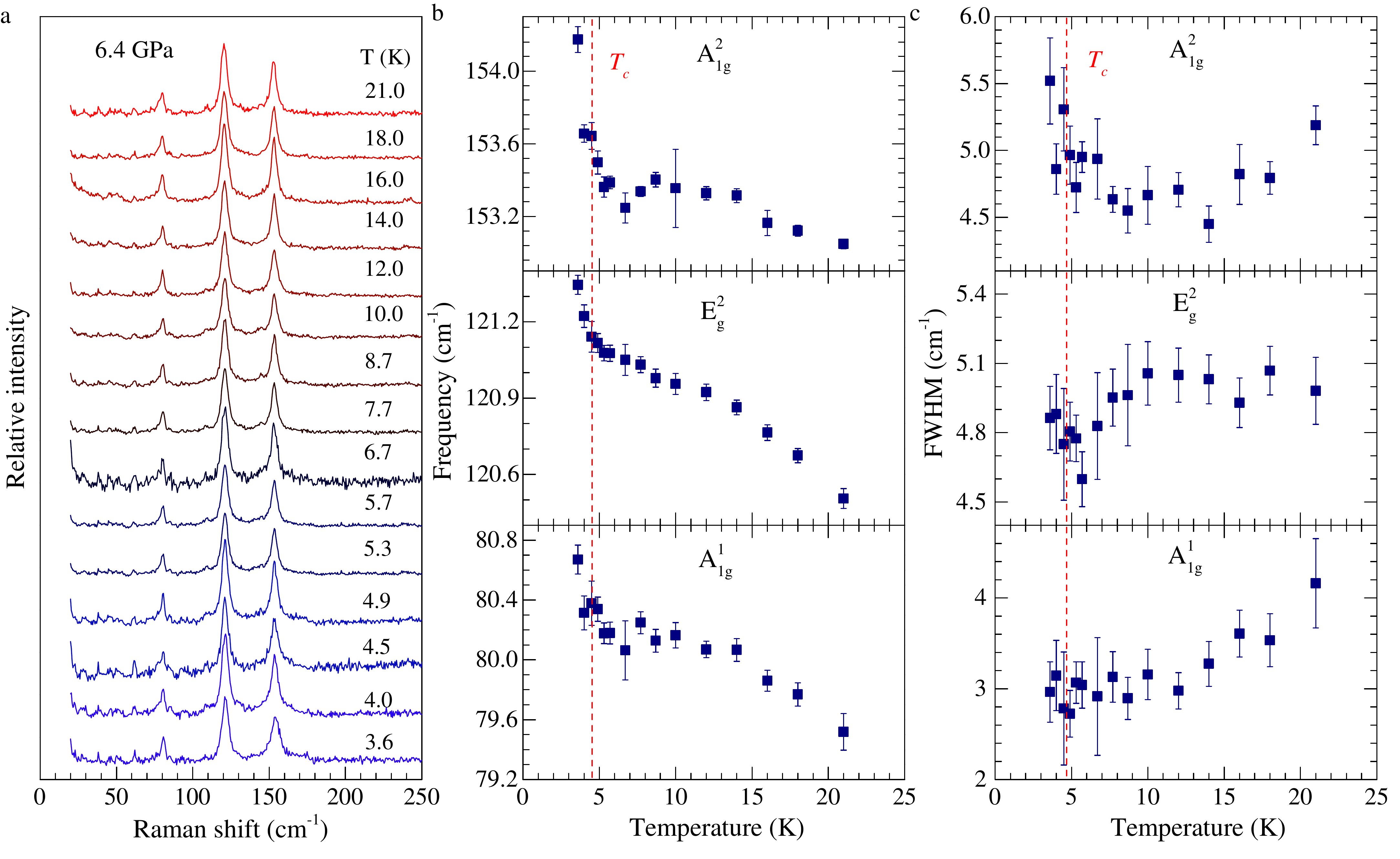}
\end{center}
\vspace{-0.5cm}
\noindent\textbf{Extended Data Fig. 6 $\mid$ Temperature-dependent Raman spectra for Bi$_{2}$Te$_{2.7}$Se$_{0.3}$ across the superconducting transition in Phase I at pressure of 6.4 GPa}. \textbf{a}, Raman spectra at low temperatures down to 3.6 K. \textbf{b} and \textbf{c}, Temperature dependence of the frequency and FWHM of the $A_{1g}^{1}$, $E_{g}^{2}$, and $A_{1g}^{2}$ mode. The $T_{c}$ is indicated by the dashed line.

\end{document}